\documentclass[]{aa}
\usepackage{psfig}
\begin{document}
\title{Lithium in stars with exoplanets\thanks{Based on
    observations collected at the La Silla Observatory, ESO (Chile),
    with the {\footnotesize CORALIE} spectrograph at the 1.2 m
    Euler Swiss telescope, and with the {\footnotesize FEROS} spectrograph
    at the 1.52 m ESO telescope, and using the UES
    spectrograph at the 4.2 m William Herschel Telescope (WHT) and SARG
    spectrograph at the 3.5 m Telescopio Nazional Galileo
    on La Palma (Canary Islands).}}

\author{G.~Israelian\inst{1}
\and  N.~C.~Santos\inst{2,3}
\and  M.~Mayor\inst{3}
\and  R.~Rebolo\inst{1}
}
\offprints{G.Israelian (gil@iac.es)}
\institute{Instituto de Astrof{\'\i}sica de Canarias,, E-38205 La Laguna,
Tenerife, Spain
\and
Centro de Astronomia e Astrof{\'\i}sica da Universidade de Lisboa,
        Observat\'orio Astron\'omico de Lisboa, Tapada da Ajuda, 1349-018
        Lisboa, Portugal
\and
Observatoire de Gen\`eve, 51 ch.\ des
Maillettes, CH--1290 Sauverny, Switzerland
}
\date{Received; accepted}

\titlerunning{Lithium in stars with exoplanets}

\abstract{
We present a comparison of the lithium abundances of stars with
and without planetary-mass companions. New lithium abundances
are reported in 79 planet hosts and 38 stars from a comparison sample.
When the Li abundances of planet host stars are compared with the 157 stars
in the sample of field stars of Chen et al.\ (2001) we find that the Li abundance
distribution is significantly different, and that there is a possible excess of
Li depletion in planet host stars with effective temperatures in the range 
5600--5850 K, whereas
we find no significant differences in the temperature range 5850--6350 K.
 We have searched for  statistically significant correlations between the Li abundance
of parent stars and various parameters of the planetary companions. We do not
find any  strong correlation, although there are may be a hint
of a possible gap in the Li distribution of massive planet host stars.

\keywords{stars: abundances -- stars: chemically peculiar -- planetary systems}
}

\maketitle

\section{Introduction}

The extrasolar planetary systems detected to date are probably not a representative
sample of all planetary systems in the Galaxy. Indeed, the detection of a giant
planet with a mass $M_{\rm p}\sin i$=0.47 $M_{\rm J}$ (Jupiter masses) orbiting the
solar-type star 51 Peg at 0.05 AU (Mayor \& Queloz 1995) was not anticipated.
The Doppler method, which formed the basis of the discovery of more than 100 extrasolar planets,
is clearly biased, being most sensitive to massive planets orbiting close to their
parent stars. These surveys have established that at least  $\sim$7\% of solar-type
stars host planets (Udry \& Mayor 2001). On the other hand, we can learn a
lot about the formation
and evolution of planetary systems by studying in detail properties of stars with planets.
Although extrasolar planetary systems differ from the Solar System, the host
stars themselves do not appear to be distinguished by their kinematic or physical
properties. They are normal main sequence stars that are metal-rich relative to nearby
field stars (Gonzalez 1998; Santos, Israelian \& Mayor 2000, 2001;
Gonzalez et al.\ 2001; Santos et al.\ 2003a). Possible explanations for the high
metallicities of the stars with exoplanets involve primordial effects
(Santos et al. 2001, 2003a; Pinsonneault, DePoy \& Coffee 2001) and the ingestion
of rocky material, planetesimals and/or metal-rich gaseous giant planets
(Gonzalez 1998; Gonzalez et al.\ 2001; Laughlin \& Adams \cite{La97};
Murray et al.\ \cite{Mu01a}; Murray \& Chaboyer 2002).
While our recent discovery (Israelian et al.\ 2001, 2003)  of a significant amount of
$^6$Li in the planet host HD\,82943 clearly suggests that the accretion of planetesimals
or maybe entire planets has indeed taken place in some stars, we cannot state that this effect
is responsible for the metallicity enhancement in planet-harbouring stars. This question
can possibly be answered if we analise the abundances of Li, Be (beryllium) and the isotopic
ratio $^6$Li/$^7$Li in a large number of planet-bearing stars. Combined with precise
abundance analyses of Fe and other elements, these studies may even allow us to distinguish
between different planet formation theories (Sandquist et al.\ 2002).
The light elements Li and Be are very important tracers of the internal structure and
pre-main sequence evolution of solar type stars. In some way, studies of Be and Li
complement each other. Lithium is depleted at much lower temperatures (about 2.5 milion K)
than Be. Thus, by measuring Li in stars where Be is not
depleted (early G and late F) and Be in stars where Li is depleted (late G and K)
we can obtain crucial information about the mixing, diffusion and angular momentum history
of exoplanet hosts (Santos et al.\ 2002).

Gonzalez \& Laws (2000) presented a direct comparison of Li abundances
among planet-harbouring stars with field stars and proposed that the former have less Li.
However, in a critical analysis of this problem Ryan (2000) concludes that planet hosts
and field stars have similar Li abundances. Given the large number of 
planet-harbouring stars
discovered to date, we have decided to investigate the Li problem  and look for
various statistical trends. We have attempted to remove and/or minimize
any bias in our analysis following the same philosophy as Santos et al.\ (2001).
Here, we present the results of Li analyses in 79 stars with planets and 38 stars
from a comparison sample consisting of stars without detected planets from a CORALIE
sample (Santos et al.\ 2001). Comparison of Li abundance in planet hosts and a sample of
157 solar-type stars from Chen et al.\ (2001) is presented and different physical processes
that can affect the evolution of the surface abundance of Li in stars with exoplanets
are discussed.

\section{Observations and analysis}

The spectroscopic observations of our targets were carried out during different runs
using the 4.2 m WHT/UES (La Palma), the 3.5 m TNG/SARG (La Palma), the 1.52 m
ESO (La Silla) and the 1.2 m Swiss/CORALIE (La Silla).  The same data were used in 
recent papers by Santos et al.\ (2003a) and Bodaghee et al.\ (2003). Observations
with the WHT/UES were obtained using the E31 grating and
a 1.1 arcsec slit providing a resolving power 55\,000. The TNG observations were carried out
with the SARG spectrograph and a two EEV CCD mosaic of $4096\times 2048$
pixels of  size  $15\ \mu {\rm m} \times 15\ \mu{rm m}$. Resolving power $\sim$57\,000 
was achieved with 1 arcsec slit.
All the WHT and TNG images were reduced using standard IRAF routines. Normalized
flats created for each observing night were used to correct the pixel-to-pixel
variations and a ThAr lamp was used to find a dispersion solution.
The ESO 1.52 m/FEROS (La Silla, Chile) observations were carried out using
 two EEV  detector mosaic of $4096\times 2048$  pixels (size 15 $\mu {\rm m} \times 15\ \mu{\rm m} $).
Automatic spectral reduction was carried out using special FEROS software.
In the present analysis we used the same spectral synthesis tools as in
Santos et al.\ (2001, 2002, 2003a) and Israelian et al.\ (2001, 2003).
The stellar parameters (Tables 2 and 3) were taken from Santos et al.\ (2003a) and
Bodaghee et al.\ (2003). The orbital parameters of planets were obtained from
the Extrasolar Planets Encyclopaedia (http://www.obspm.fr/encycl/encycl.html)
compiled by Jean Schneider.

\section{Lithium in solar-type stars}

The light element Li provides information regarding the redistribution and mixing of
matter within a star. Standard evolutionary models predict that the Li abundance 
in main sequence stars should depend uniquely on the stellar effective temperature,
age (chromospheric activity) and metallicity (see for example D'Antona F. \& Mazzitelli 1994).
Mass (or $T_\mathrm{eff}$) is the first parameter that governs the Li depletion in
solar-type stars. Age is the second parameter which accounts for a MS
(main sequence) depletion and is also linked with chromospheric activity.
A third parameter(or perhaps parameters) might be a metallicity and/or rotation. 
This is confirmed by the
analysis of the correlation matrix for the parameters governing the surface abundance
of Li (Pasquini et al.\ 1994). On the other hand, we know that classical models of
stellar evolution neglect several important physical processes that
are important for interpreting the photospheric Li abundance in solar-type stars.
Gravitational settling (downward motions), thermal diffusion (downward
motions) and radiative acceleration (upward) are among the most
important. Main sequence mass loss and slow mixing via gravity waves
(Garc\'\i a L\'opez \& Spruit 1991; Montalb\'an \& Shatzman 2000)
and rotation via angular momentum loss (e.g. Pinsonneault et al.\ 1990;
Vauclair et al.\ 1978; Maeder 1995; Zahn 1992) make the physics of depletion
even more complicated.

For solar-type stars, two important observational facts need to be explained: the high
dispersion in Li abundance for stars of similar temperature, age and metallicity in
open clusters (Pasquini et al.\ 1994) and the large Li deficiency in the Sun. On the other hand,
observations indicate that rapidly rotating stars preserve more Li than slow rotators
of the same mass (Randich et al.\ 1997; Stauffer et al.\ 1997  Garc\'\ i a L\'opez et al.\ 1994).
However, this is not enough to explain
the large Li scatter since several Li rich stars in the Pleiades are slow rotators
(King et al.\ 2000). It has also been shown that tidally locked binaries in the Hyades have
much higher Li abundances than single stars in the same cluster (Thorburn et al.\ 1993;
Deliyannis et al.\ 1994). Nevertheless, Ryan \& Deliyannis (1995) found close
binaries in Pleiades with normal Li abundances, but, given the young age of the
cluster, this may not be conclusive. Numerous observations strongly indicate
that there must be an additional parameter, or parameters, to control the surface 
abundance of Li in solar-type stars.

Lithium destruction is sensitive to the detailed chemical composition of the
stellar matter. Depletion of Li anticorrelates with helium and deuterium content 
because of opacity effects. The increase of metal opacities in solar-type stars is responsible for
the transition between radiative and convective energy transport. The main contributors to
the total opacity at BCZ (base of the convection zone) of the present Sun are oxygen
(36 \%) and iron (20\%) (see Table 3 of Piau \& Turck-Chieze 2002). However, 
observations show no clear correlation between
Li and [Fe/H] in the metallicity range $- 1  <$ [Fe/H] $<$ 0.2 (e.g. Pasquini et al.\ 1994;
Chen et al.\ 2001) and Li and [O/H] in the range $-0.5 <$ [O/H] $<$ 0.4 (Pomp\'eia et al.\ 2002).

Li depletion already takes place in the pre-main sequence (PMS) phase of stellar
evolution and increases with decreasing stellar mass. During the PMS, stars slowly contract
towards the zero-age main sequence (ZAMS) in quasi-hydrostatic equilibrium within the Kelvin--Helmoltz
timescale. The PMS lifetime varies from 30 to 100 Myr for stars with 1.4 and 0.8 $M_{\sun}$,
respectively. The stars pass several stages of light-element burning during contraction.
Initial energy production is provided by deuterium fusion at 5 $\times$ 10$^5$ K. According
Palla \& Stahler (1991), this phase stops the contraction at radius 5--6 $R_{\sun}$ for a
1 $M_{\sun}$ star. For solar-mass stars deuterium fusion starts at the age
of $\sim$4 $\times$ 10$^4$ yr and continues for $\sim$2 $\times$ 10$^5$ yr. The Li depletion
starts 1.4 Myr before the appearance of a radiative core. The temperature at the
BCZ for a 1 $M_{\sun}$ star rapidly increases from 10$^6$ to 4 $\times$ 10$^6$ at 2 Myr and
then slowly decreases toward the value almost equal to that observed in the present Sun
(i.e.\ 1.2 $\times$ 10$^6$ K). This results in rapid burning of Li within 2--20 Myr.

Mass accretion in the T Tauri phase can affect surface abundance of Li
in several ways. First of all, it modifies the stellar mass and therefore
alters the stratification. Second, it adds  matter with ISM
abundances to the surface of the star thus modifying the chemical
composition of the atmosphere. And finally, accretion changes the boundary
conditions. Mass accretion rates in T Tauri stars vary between
10$^{-6}$ and 10$^{-8}$ $M_{\sun}$ yr$^{-1}$ (Hartigan, Edwards \& Ghandour 1995).
Global accreted mass could be of the order of few times 10$^{-2}$ $M_{\sun}$
(Hartmann 1998). During the accretion process the star depletes both its
initial Li and also the Li it receives from accretion. It is clear that
accretion will enhance surface Li abundance if it could last long
enough, i.e.\ after PMS depletion. More than 90\% of the final stellar mass
is accreted during less than 1 Myr before the classical T Tauri phase
(Andre, Ward-Thomson \& Barsony 1999). This phase is followed by
slow (but most probably variable) accretion during some 30--50 Myr.
Recent computations (Piau \& Turck-Chieze 2002) suggest that the more
the star accretes the more it depletes Li because of the dominant structural
effects. Apparently, accreted Li does not compensate for the additive
burning because of the lower mass of the star. Accretion rates as low
as 10$^{-9}$ $M_{\sun}$ (or even lower) are required in order to counteract
the mass effect.  Let us also note that internal rotation on the PMS
also has an important effect on Li as the core and surface may have different
rotation rates.

The existence of strong Li depletion in the Sun is inconsistent with
classical models. In order to explain the observations, lithium must be
transported from the convection zone to the hot layers where the
temperature is more than 2.5 $\times$ 10$^{6}$ K. However, the overall
effect must be small in order to
allow for some Li preservation in the solar atmosphere after 4.5 Gyr of MS
evolution. The real problem is how the Li nuclei can cross the gap between the hot
layers and the BCZ. Overshooting convection (Ahrens et al.\ 1992) and
anisotropic turbulence stabilized by the radial temperature gradient
(Zahn 1992) are among the mechanisms most commonly discussed in the
literature. This transport is less effective in rapidly rotating stars.
The amount of Li depletion in the Sun cannot be explained by rotation
and convective diffusion since the timescales of these processes are
12 days (Noyes et al.\ 1984) and 100 yr (R\"{u}diger \& Pipin 2001), respectively.
This clearly indicates that any non-convective mixing must be very slow.

The presence of a large ($\sim$1 dex) Li gap in solar-type stars with
5600  K $< T_\mathrm{eff} <$ 5900 K
has been suggested by different authors  (see for example Pasquini et al.\ 1994;
Chen et al.\ 2001). The Sun belongs to the group with low Li abundance with
$\log \epsilon(\rm Li)$ = 1.16 (M\"{u}ller, Peytremann \& De La Reza 1975) and according
Pasquini et al.\ (1994), about 50\% of these stars having similar $T_\mathrm{eff}$
and age as the Sun have suffered an equally severe Li depletion during their MS lifetime.
Main-sequence depletion appears to be a slow and more complicated process.

In summary, a large spread of Li abundance exists in solar-type stars of similar
age, mass and metallicity. This spread cannot be explained solely in terms of
these parameters. The large Li dispersion may be produced during MS evolution by a
still an unknown mechanism. Rotationally induced mixing and MS mass-loss could produce
different Li abundance in stars with similar mass, age and chemical composition.
What is not clear is why these non-standard mixing processes produce a "gap" on the
Li morphology for stars with 5600 K $< T_\mathrm{eff}$ $<$ 5900 K but not
a large scatter.

\section{Evolution of Li In Stars With Exoplanets}

\subsection{Accretion of planets and planetesimals}

A large  number of comets that plunge into the Sun have been discovered by
SOHO (Raymond et al.\ 1998). There is almost no doubt that the flux of resonant
asteroids that strike the Earth and the Sun was much higher in the past. The sweeping of
mean motion resonances was caused by a dissipation of a protoplanetary gas disc
and the migration of Jupiter and Saturn to their current positions. These processes
led to the depletion of the outer belt and the accretion of rocky matter on to
the Sun. The belt between Earth and Jupiter was more massive in the past as is 

evidenced by the interpolation of the surface density of iron material in the
Solar System planets (Weidenschilling 1977). Other independent evidence comes
from the accretion of the asteroids over short time scales as indicated by the
analysis of meteorites (Wetherill 1989). It is believed that up to 5 Earth masses
would have been between Mars and Jupiter, about half of which have accreted in
the Sun.

\begin{figure}
\psfig{figure= 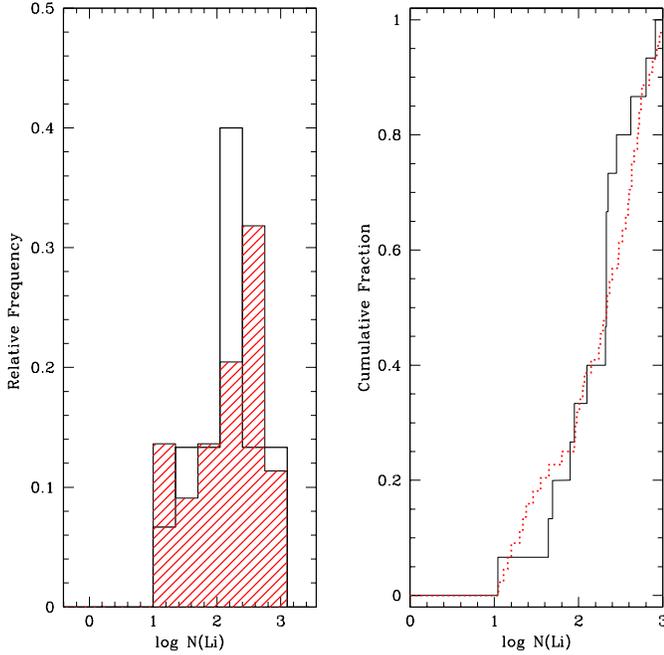,width=9.5cm,height=9.5cm,angle=360}
\caption[]{Lithium distribution for stars with planets (hatched histogram) compared with the
same distribution for the field stars (Table 2) without planets (empty histogram).
A Kolmogorov--Smirnov test shows the probability for the two populations being  part
of the same sample to be 0.6.}
\end{figure}

\begin{figure}
\psfig{figure=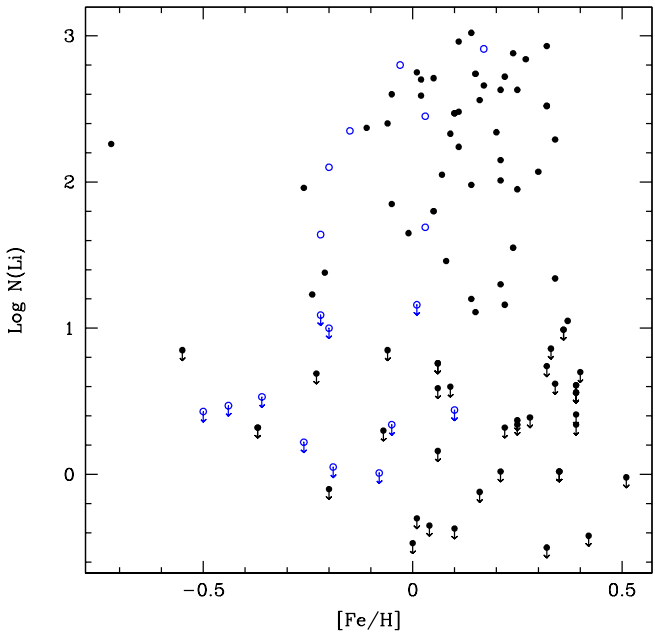,width=9.5cm,height=9.5cm,angle=360}
\caption[]{Lithium versus metallicity for stars with (filled dots) and without
(empty circles) planets from Santos et al.\ (2001).}
\end{figure}

Slow accretion of  planetesimals was invoked in order to explain the [Fe/H]
distribution of planet-harbouring stars. Based on an analysis of 640 solar-type
stars Murray et al.\ (2001) have suggested that the main sequence accretion of a
chondritic matter is a common process in MS stars. These authors have proposed
that most, if not all, solar-type stars accreted 0.4 Earth masses of iron after
they reached the main sequence. In a different paper Murray \& Chaboyer (2002)
conclude that an average of 6.5 Earth masses of iron must be added to the 
planet-harbouring stars in order to explain the mass--metallicity and age--metallicity relations.
Given that a small fraction of protostellar discs have masses around 0.1\ $M_{\sun}$,
such discs would contain at least 10 Earth masses of iron even if their metallicity is
[Fe/H] = $-$0.5. It is of course not clear which fraction of planetesimals will be accreted
in stars with different atmospheric parameters or when. But in principle, one can be
sure that there is a large amount of iron available in protoplanetary discs
in the form of planetary embryos, asteroids and planetesimals. In some
planetary systems, this matter may be accreted during MS evolution making the parent
stars metal rich. Observational biases and poorly known convection zone  masses
of stars with $M>$ 1.2 \ $M_{\sun}$ are responsible for the current debate on
the source of metal enrichment in planet host stars (Santos et al.\ 2001, 2003a;
Pinsonault et al.\ 2001; Murray \& Chaboyer 2002).

Accretion of a few Earth masses of planetesimals during early MS evolution will
strongly modify $^7$Li abundances in these stars. Moreover, in stars with
$T_\mathrm{eff} >$ 5900 K a large amount of the added $^6$Li
 may avoid  destruction via mixing given the depth of the convection zone
(Montalb\'an \& Rebolo 2002). Following the estimates of Murray et al.\ (1998, 2001)
and Murray \& Chaboyer (2002), we would expect a large amount of $^6$Li in the
atmospheres of late F/early G main sequence metal-rich planet hosts. Our detection
of $^6$Li in HD\,82943 (Israelian et al.\ 2001, 2003) certainly suggests that
this test should be continued in other systems.

Numerical simulations of inward migration suggest that planets
may be ingested in some systems. Different physical mechanisms may lead to 
planet engulfment and each of them have their characteristic timescales.
Classical migration caused by tidal interaction (Lin et al.\ 1996) operates on
short time scales (a few Myr) and will add planetary Li to the star
when the latter is still evolving towards the main sequence. This may
not have a large affect on the surface abundance of $^7$Li and $^6$Li
since these nuclei will be destroyed in hot stellar interiors owing to the
efficient convection. The time scale of planet accretion brought about by multi-body
interactions may be much longer (up to 100 Myr, Levison, Lissauer \& Duncan 1998)
compared with the pre-MS evolution lifetime; therefore, this process may
modify surface abundances of both Li isotopes. Dynamical friction is another
possibility for the accretion of a large amount of rocky matter during several
hundreds  of Myr or even Gyr.

We conclude that there are various physical process that may lead
to the accretion of matter by stars with extrasolar planets during
their MS lifetime. These processes will modify surface abundance of Li.

\subsection{Stellar Activity Caused by Interaction With Exoplanets}

It is well know that stellar chromospheric or coronal activity increases when
two stars interact with each other (e.g.\ RS CVn systems). This effect is mostly
caused by enhanced dynamo activity brought about by rotational synchronization
and spin-up. Activity can also be triggered by tidal effects (Catalano et al.\ 1996).
Resulting flares may be a source of Li just as it is produced in the Sun (Livshits 1997).
The effects of tidal and magnetic interaction are also expected to occur in stars
with exoplanets. These effects have recently been considered by Cuntz, Saar \&
Musielak (2000). We also note that Shkolnik, Walker \& Bohlender (2003) have detected the
synchronous enhancement of Ca{\sc ii} H \& K emission lines with the short period
planetary orbit in HD\,179949. Another example of the stellar activity triggered by 
a star-planet interaction was presented by Santos et al. (2003b) in HD\,192263.

Present exoplanet surveys are targeting old, chromospherically inactive,
slowly rotating stars. This observational bias does not allow us to discover
any possible relationship between rotation, chromospheric activity and Li in
planet-harbouring stars.  The reason for avoiding young and active stars lies in
surface spots, which introduce systematic variations in the Doppler velocities of
stellar absorption lines. While solar flares produce Li in spallation
reactions, the amount of Li and the dynamics of flares are such that no Li
atoms are accreted in the stellar photosphere (Ramaty et al.\ 2000).
However, the so called {\it superflares}, if they exist, may modify the surface
abundance of Li in cool stars and planet hosts, in particular. There
have been nine observations of old solar-type stars indicating very strong
flares with durations from minutes to days (Rubenstein \& Shaefer 2000;
Schaefer, King \& Deliyannis 2000). There is no clear theoretical
interpretation of these events while the link with hot jupiters
has already been put forward (Rubenstein \& Shaefer 2000). Strong magnetic
fields of short period giant planets may become entangled with the magnetic
fields of their parent stars and release large amounts of energy in
superflares via magnetic reconnection. The amount of energy created
in these flares is large (10$^{33}$--10$^{38}$ erg) enough to create
a substantial amount of Li (Livshits 1997; Ramaty et al.\ 2000).

If strong flares are able to enhance atmospheric Li in planet hosts, then
we may expect parent stars in short period systems to have more Li on average.
Such flares will create not only $^7$Li but also $^6$Li.

\subsection{The Tidal Effects In Short-Period Systems}

Engulfment of planets and brown dwarfs has been suggested as the cause
of the high rotational velocities in some field red giants (Stefanik et al.\ 2001)
and blue horizontal branch stars (Peterson, Tarbell \& Carney 1983; Soker \& Harpaz
2000). A theoretical examination of the effects of planet engulfment on 
angular momentum evolution and mass loss rates from giants was recently carried out
by Livio \& Soker (2002).

Various observations confirm a correlation between the lithium content and the
angular momentum lost by solar-type stars (Garc\'\i a L\'opez et al.\ 1994;
Randich et al.\ 1998). The physics of this relationship was
explored  by different authors (Pinsonneault et al.\ 1990; Zahn 1992, 1994).
A link between lithium depletion and angular momentum loss is also predicted
for binary systems. Viscous dissipation of time-dependent tidal effects may produce
the circularization of the binary system orbit and synchronization between stellar
rotation and orbital motion (Zahn 1977). While single stars spin down because of
angular momentum lost via stellar winds, stars in binary systems may spin up
as a result of the momentum gained from  orbital migration. Many observations
(De Medeiros, Do Nascimento \& Mayor 1997; Costa et al.\ 2002.)
show that stars in binary systems with a period less than the critical period
for synchronization generally have enhanced rotation compared with their single
counterparts. There are strong indications that lithium is less depleted in
short-period binary systems with enhanced rotation.

It is well known that short-period planets have tidal interactions with
their parent stars. If a star's rotation period is greater than that of the planets,
the star will spin up because of tidal friction. This may prevent strong Li
depletion. Momentum conservation
will lead to a decrease in the semimajor axis of the planet's orbit. This interaction
was invoked in order to explain the absence of massive planets at $a < $0.1 AU
(P\"{a}tzold \& Rauer 2002). A critical test for this scenario would be a comparison
of rotational velocities of stars in short and long period planetary systems.
However, this may not be easy given the complex and time-dependent nature of
the core--envelope evolution and the star--planet interaction. The orbital angular
momentum of the close-in planets transferred to the star may influence the angular
momentum evolution of the remaining planets in the system. The rotationally
decoupled convective layer  may spin up and force the remaining planets
to spiral outward. The enhanced angular momentum of the convective layer may create a
large shear instability at the interface between the convective and radiative zones
that may result in mixing between the convection zone and stellar interior by a
decreasing surface abundance of Li. Planetary migration and/or consumption may also
enhance magnetic activity via the dynamo effect. Consequently, the star will spin down
because of enhanced magnetic breaking.

The angular momentum history of solar-type stars is strongly influenced by the formation and
evolution of planetary systems. The wide dispersion in rotation rates of cluster stars has been
explained (Edwards et al.\ 1993) by invoking disk interactions in the pre-MS. This
phenomenon, as well as the formation of planets, may prevent some stars from ever
passing through a fast rotator phase near the ZAMS. It is believed that magnetic interactions
between pre-MS stars and their discs, and  the formation of planetary systems with different
characteristics, create a wide range of initial rotation periods that virtually converge
on the main sequence. Thus, stars with similar age, mass and metallicity may arrive on the
MS with similar rotation velocities but different amounts of Li.

 Barnes (2003) has recently proposed that rotating solar-type stars lie primarily on two
sequences. Stars evolve from a core--envelope decoupled state to a coupled state. It
is interesting to investigate whether the physics behind the two rotational sequences
of Barnes (2003) has anything to do with the Li gap of Chen et al.\ (2001) and
Pasquini et al.\ (1994 ). The planetary migration may also leave their
signatures on period--colour diagrams of clusters and field stars.

\begin{figure}
\psfig{figure=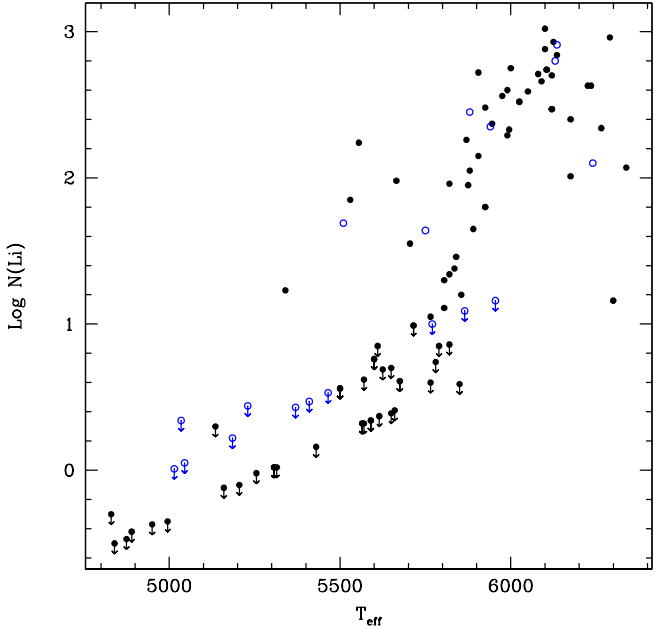,width=9.5cm,height=9.5cm,angle=360}
\caption[]{Lithium versus effective temperature for stars with (filled dots) and
without planets (empty circles) from Santos et al.\ (2001).}
\end{figure}

\begin{figure}
\psfig{figure=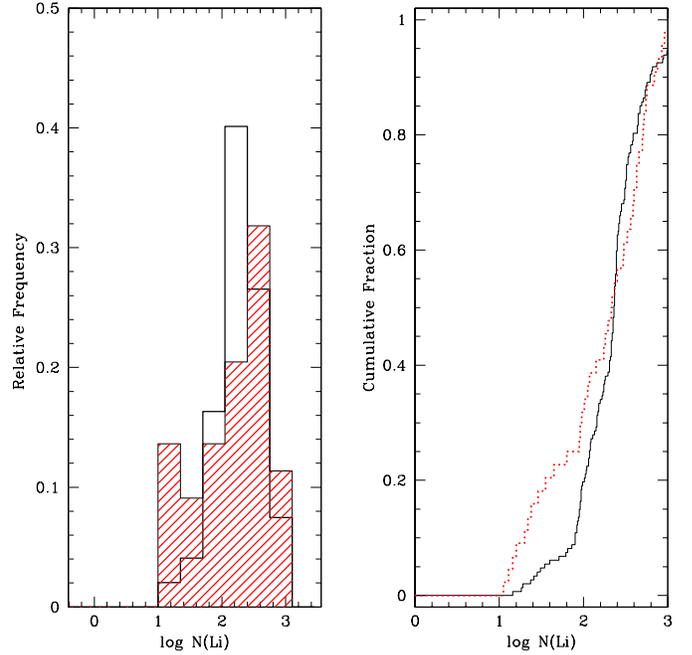,width=9.5cm,height=9.5cm,angle=360}
\caption[]{Lithium distribution for stars with planets (hatched histogram) compared
with the same distribution for the field stars from Chen et al.\ 2001. (empty histogram).
A Kolmogorov--Smirnov test shows the probability for the two populations being a part of
the same sample to be 0.2.}
\end{figure}

\begin{figure}
\psfig{figure=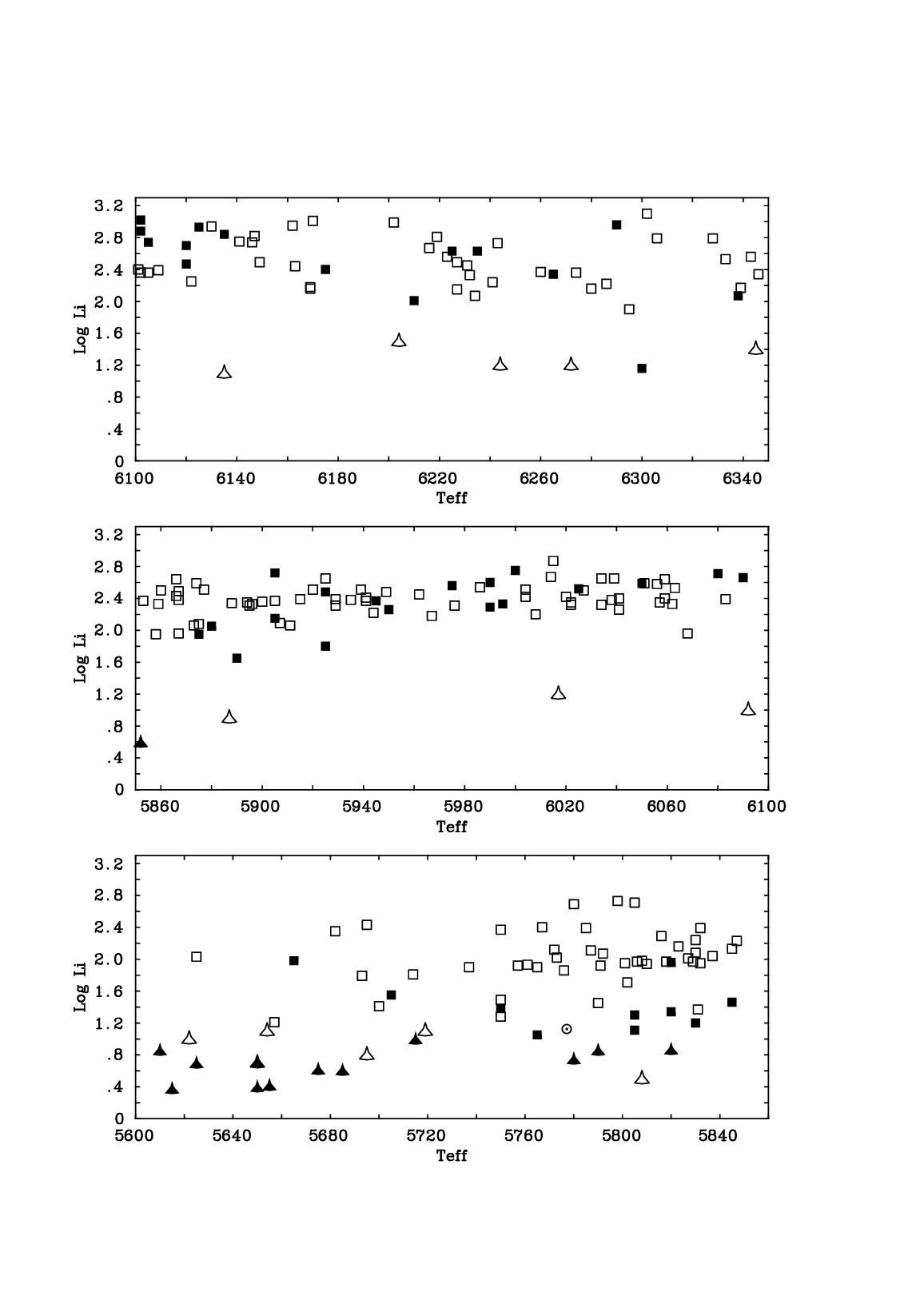,width=10.5cm,height=16.5cm,angle=360}
\caption[]{Lithium versus effective temperature for stars with planets
(filled squares) and the comparison sample of Chen et al.\ (empty squares). Upper
limits are filled (planet hosts) and empty (comparison sample) triangles.
The position of the Sun is indicated.}
\end{figure}

\begin{figure}
\psfig{figure=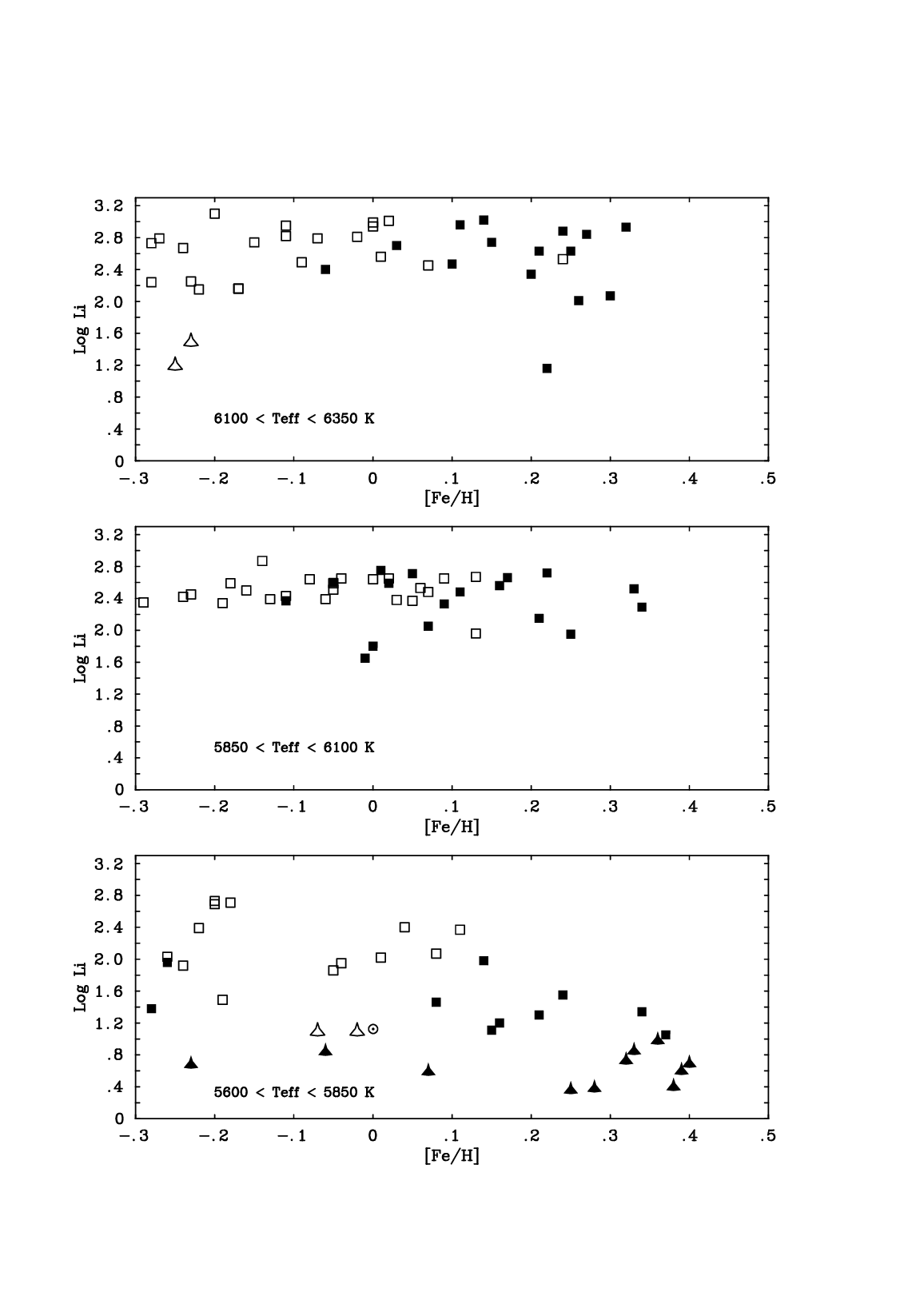,width=10.5cm,height=16.5cm,angle=360}
\caption[]{Lithium versus metallicity for stars with planets
(filled squares) and the comparison sample of Chen et al.\ (empty squares) for three
regions of effective temperature. Upper limits are filled (planet hosts) and
empty (comparison sample) triangles. The position of the Sun is indicated.}
\end{figure}

\def\baselinestretch{1}
\begin{table*}[t]
\caption[]{Determined atmospheric parameters and Li abundances for a set of stars
with planets and brown dwarf companions}
\begin{scriptsize}
\begin{tabular}{lccccl}
\hline
\noalign{\smallskip}
Star & $T_\mathrm{eff}$ & $\log{g}$       & $\xi_t$    & [Fe/H]  & $\log \epsilon({\rm Li}$) \\
     & (K)              &  (cm\,s$^{-2}$) & (km\,s$^{-1}$) &       &   \\
\hline \\
\object{HD\,142 }   & 6290 & 4.38 & 1.91   & 0.11     & 2.96        \\
\object{HD\,1237  } & 5555 & 4.65 & 1.50   & 0.11     & 2.24        \\
\object{HD\,2039  } & 5990 & 4.56 & 1.23   & 0.34     & 2.29       \\
\object{HD\,4203  } & 5650 & 4.38 & 1.15   & 0.40     & $<$0.70      \\
\object{HD\,4208  } & 5625 & 4.54 & 0.95   & $-$0.23  & $<$0.69      \\
\object{HD\,6434  } & 5790 & 4.56 & 1.40   & $-$0.55  & $<$0.85      \\
\object{HD\,8574  } & 6080 & 4.41 & 1.25   & 0.05     & 2.71        \\
\object{HD\,9826  } & 6120 & 4.07 & 1.50   & 0.10     & 2.47        \\
\object{HD\,10697 } & 5665 & 4.18 & 1.19   & 0.14     & 1.98      \\
\object{HD\,12661 } & 5715 & 4.49 & 1.09   & 0.36     & $<$0.99    \\
\object{HD\,13445 } & 5190 & 4.71 & 0.78   & $-$0.20  & $<-$0.1    \\
\object{HD\,16141 } & 5805 & 4.28 & 1.37   & 0.15     & 1.11       \\
\object{HD\,17051 } & 6225 & 4.65 & 1.20   & 0.25     & 2.63       \\
\object{HD\,19994 } & 6210 & 4.20 & 1.52   & 0.26     & 2.01       \\
\object{HD\,20367 } & 6100 & 4.55 & 1.31   & 0.14     & 3.02       \\
\object{HD\,22049 } & 5135 & 4.70 & 1.14   & $-$0.07  & $<$0.3    \\
\object{HD\,23079 } & 5945 & 4.44 & 1.21   & $-$0.11  & 2.37       \\
\object{HD\,23596 } & 6125 & 4.29 & 1.32   & 0.32     & 2.93       \\
\object{HD\,27442 } & 4890 & 3.89 & 1.24   & 0.42     & $< -$0.42 \\
\object{HD\,28185 } & 5705 & 4.59 & 1.09   & 0.24     &  1.55    \\
\object{HD\,30177 } & 5590 & 4.45 & 1.07   & 0.39     & $<$0.35   \\
\object{HD\,33636 } & 5990 & 4.68 & 1.22   & $-$0.05  & 2.60      \\
\object{HD\,37124 } & 5565 & 4.62 & 0.90   & $-$0.37  & $<$0.32   \\
\object{HD\,38529 } & 5675 & 4.01 & 1.39   & 0.39     & $<$0.61  \\
\object{HD\,39091 } & 5995 & 4.48 & 1.30   & 0.09     & 2.33      \\
\object{HD\,46375 } & 5315 & 4.54 & 1.11   & 0.21     & $<$0.02  \\
\object{HD\,50554 } & 6050 & 4.59 & 1.19   & 0.02     & 2.59     \\
\object{HD\,52265 } & 6100 & 4.29 & 1.31   & 0.24     & 2.88      \\
\object{HD\,74156 } & 6105 & 4.40 & 1.36   & 0.15     & 2.74      \\
\object{HD\,75289 } & 6135 & 4.43 & 1.50   & 0.27     & 2.84     \\
\object{HD\,75732}  & 5307 & 4.58 & 1.06   & 0.35    & $<$0.02   \\
\object{HD\,80606}  & 5570 & 4.56 & 1.11   & 0.34     & $<$0.62  \\
\object{HD\,82943}  & 6025 & 4.54 & 1.10   & 0.33     & 2.52      \\
\object{HD\,83443}  & 5500 & 4.50 & 1.12   & 0.39     & $<$0.56  \\
\object{HD\,89744}  & 6338 & 4.17 & 1.55   & 0.30     & 2.07      \\
\object{HD\,92788}  & 5820 & 4.60 & 1.11   & 0.34     & 1.34      \\
\object{HD\,95128}  & 5925 & 4.45 & 1.24   & 0.00     & 1.8       \\
\object{HD\,106252} & 5890 & 4.40 & 1.06   & $-$0.01  & 1.65     \\
\object{HD\,108147} & 6265 & 4.59 & 1.40   & 0.20     & 2.34     \\
\object{HD\,108874} & 5615 & 4.58 & 0.93   & 0.25     & $<$0.37   \\
\object{HD\,114386} & 4875 & 4.69 & 0.63   & 0.00     & $< -$0.47  \\
\object{HD\,114729} & 5820 & 4.20 & 1.03   & $-$0.26  & 1.96       \\
\object{HD\,114762} & 5950 & 4.45 & 1.0    & $-$0.60  & 2.26      \\
\object{HD\,114783} & 5160 & 4.75 & 0.78   & 0.16     & $-$0.12    \\
\object{HD\,117176} & 5530 & 4.05 & 1.08   &$-$0.05   & 1.85      \\
\object{HD\,121504} & 6090 & 4.73 & 1.35   & 0.17     & 2.66       \\
\object{HD\,128311} & 4950 & 4.80 & 1.00   & 0.10     & $< -$0.37  \\
\object{HD\,130322} & 5430 & 4.62 & 0.92   & 0.06     & $<$0.16   \\
\object{HD\,134987} & 5780 & 4.45 & 1.06   & 0.32     & $<$0.74    \\
\object{HD\,136118} & 6175 & 4.18 & 1.61   & $-$0.06  & 2.40      \\
\object{HD\,141937} & 5925 & 4.62 & 1.16   & 0.11     & 2.48      \\
\object{HD\,143761} & 5750 & 4.10 & 1.2    &$-$0.29   & 1.38      \\
\object{HD\,145675} & 5255 & 4.40 & 0.68   & 0.51     & $< -$0.02  \\
\object{HD\,147513} & 5880 & 4.58 & 1.17   & 0.07     & 2.05       \\
\object{HD\,150706} & 6000 & 4.62 & 1.16   & 0.01     & 2.75      \\
\object{HD\,160691} & 5820 & 4.44 & 1.23   & 0.33     & $<$0.86   \\
\object{HD\,162020} & 4830 & 4.76 & 0.72   & 0.01     & $< -$0.3   \\
\object{HD\,168443} & 5600 & 4.30 & 1.18   & 0.10     & $<$0.76   \\
\object{HD\,168746} & 5610 & 4.50 & 1.02   &$-$0.06   & $<$0.85    \\
\object{HD\,169830} & 6300 & 4.04 & 1.37   & 0.22     & 1.16       \\
\object{HD\,177830} & 4840 & 3.60 & 1.18   & 0.32     & $< -$0.50 \\
\object{HD\,178911B} & 5650 & 4.65 & 0.85   & 0.28    & $<$0.39   \\
\object{HD\,179949} & 6235 & 4.41 & 1.38   & 0.21     & 2.63      \\
\object{HD\,186427} & 5685 & 4.26 & 0.80   & 0.07     & $<$0.60   \\
\object{HD\,187123} & 5830 & 4.40 & 1.00   & 0.16     & 1.20    \\
\object{HD\,190228} & 5325 & 3.95 & 1.10   &$-$0.23   & 1.23      \\
\object{HD\,190360} & 5590 & 4.48 & 1.06   & 0.25     & $<$0.34    \\
\object{HD\,192263} & 4995 & 4.76 & 0.90   & 0.04     & $< -$0.35  \\
\object{HD\,195019} & 5845 & 4.39 & 1.23   & 0.08     & 1.46      \\
\object{HD\,196050} & 5905 & 4.41 & 1.40   & 0.21     & 2.15      \\
\object{HD\,202206} & 5765 & 4.75 & 0.99   & 0.37     & 1.05     \\
\object{HD\,209458} & 6120 & 4.56 & 1.37   & 0.03     &  2.70    \\
\object{HD\,210277} & 5560 & 4.46 & 1.03   & 0.21     & $<$0.32    \\
\object{HD\,213240} & 5975 & 4.32 & 1.30   & 0.16     & 2.56     \\
\object{HD\,216435} & 5905 & 4.16 & 1.26   & 0.22     & 2.72      \\
\object{HD\,216437} & 5875 & 4.38 & 1.30   & 0.25     & 1.95      \\
\object{HD\,217014} & 5805 & 4.51 & 1.22   & 0.21     & 1.30      \\
\object{HD\,217107} & 5655 & 4.42 & 1.11   & 0.38     & $<$0.41  \\
\object{HD\,222582} & 5850 & 4.58 & 1.06   & 0.06     & $<$0.59   \\
\noalign{\smallskip}
\hline
\end{tabular}
\end{scriptsize}
\label{tab2}
\end{table*}

\def\baselinestretch{1}
\begin{table*}[t]
\caption[]{Li abundance in a volume-limited sample of stars without detected giant
planets from Santos et al.\ (2001)}
\begin{scriptsize}
\begin{tabular}{lcccrc}
\hline
\noalign{\smallskip}
Star & $T_\mathrm{eff}$ & $\log{g}$  & $\xi_t$  & [Fe/H]  & $\log \epsilon({\rm Li}$) \\
     & (k)  &  (cm\,s$^{-2}$) & (km\,s$^{-1}$) &  &    \\
\hline \\
\object{HD\,1581 }  & 5940 & 4.44 & 1.13 & $-$0.15 & 2.35     \\
\object{HD\,4391 }  & 5955 & 4.85 & 1.22 & 0.01    & $<$1.16  \\
\object{HD\,5133 }  & 5015 & 4.82 & 0.92 & $-$0.08 & $<$0.01  \\
\object{HD\,7570 }  & 6135 & 4.42 & 1.46 & 0.17    & 2.91     \\
\object{HD\,10360}  & 5045 & 4.77 & 0.89 & $-$0.19 & $<$0.05  \\
\object{HD\,10647}  & 6130 & 4.45 & 1.31 & $-$0.03 & 2.80     \\
\object{HD\,10700}  & 5370 & 4.70 & 1.01 & $-$0.50 & $<$0.43  \\
\object{HD\,14412}  & 5410 & 4.70 & 1.01 & $-$0.44 & $<$0.47  \\
\object{HD\,20010}  & 6240 & 4.27 & 2.23 & $-$0.20 & 2.10     \\
\object{HD\,20766}  & 5770 & 4.68 & 1.24 & $-$0.20 & $<$1.00  \\
\object{HD\,20794}  & 5465 & 4.62 & 1.04 & $-$0.36 & $<$0.53  \\
\object{HD\,20807}  & 5865 & 4.59 & 1.28 & $-$0.22 & $<$1.09  \\
\object{HD\,23356}  & 5035 & 4.73 & 0.96 & $-$0.05 & $<$0.34  \\
\object{HD\,23484}  & 5230 & 4.62 & 1.13 & 0.10    & $<$0.44  \\
\object{HD\,26965A} & 5185 & 4.73 & 0.75 & $-$0.26 & $<$0.22  \\
\object{HD\,30495 } & 5880 & 4.67 & 1.29 & 0.03    & 2.45     \\
\object{HD\,36435 } & 5510 & 4.78 & 1.15 & 0.03    & 1.69     \\
\object{HD\,38858 } & 5750 & 4.56 & 1.22 & $-$0.22 & 1.64    \\
\object{HD\,40307 } & 4925 & 4.57 & 0.79 &$-$0.25  & $< -$0.09 \\
\object{HD\,43162 } & 5630 & 4.57 & 1.36 &$-$0.02  & 2.33     \\
\object{HD\,43834 } & 5620 & 4.56 & 1.10 & 0.12    & 2.32     \\
\object{HD\,50281A} & 4790 & 4.75 & 0.85 & 0.07    & $< -$0.27 \\
\object{HD\,53705 } & 5810 & 4.40 & 1.18 & $-$0.19 & 1.04     \\
\object{HD\,53706 } & 5315 & 4.50 & 0.90 & $-$0.22 & $<$0.24  \\
\object{HD\,65907A} & 5940 & 4.56 & 1.19 & $-$0.29 & $<$0.98  \\
\object{HD\,69830 } & 5455 & 4.56 & 0.98 & 0.00    & $<$0.51  \\
\object{HD\,72673 } & 5290 & 4.68 & 0.81 & $-$0.33 & $<$0.52  \\
\object{HD\,76151 } & 5825 & 4.62 & 1.08 & 0.15    & 1.90     \\
\object{HD\,84117 } & 6140 & 4.35 & 1.38 & $-$0.04 & 2.62     \\
\object{HD\,189567} & 5750 & 4.57 & 1.21 & $-$0.23 & $<$0.81  \\
\object{HD\,191408A}& 5025 & 4.62 & 0.74 & $-$0.51 & $<$0.13  \\
\object{HD\,192310 }& 5125 & 4.63 & 0.88 & 0.05    & $<$0.24  \\
\object{HD\,196761 }& 5460 & 4.62 & 1.00 & $-$0.27 & $<$0.70  \\
\object{HD\,207129 }& 5910 & 4.53 & 1.21 & $-$0.01 & 2.33     \\
\object{HD\,209100 }& 4700 & 4.68 & 0.60 & 0.01    & $< -$0.39 \\
\object{HD\,211415 }& 5925 & 4.65 & 1.27 & $-$0.16 & 1.95     \\
\object{HD\,222237 }& 4770 & 4.79 & 0.35 & $-$0.22 & $< -$0.26 \\
\object{HD\,222335 }& 5310 & 4.64 & 0.97 & $-$0.10 & $<$0.35  \\
\noalign{\smallskip}
\hline
\end{tabular}
\end{scriptsize}
\label{tab1}
\end{table*}
\def\baselinestretch{2}

\section{Correlation with stellar parameters}

\subsection{Comparison sample of Santos et al.\ (2001)}

A first look at the  of Li abundances in stars with and without
exoplanets (Tables 1 and 2) from Santos et al.\ (2001) suggests that both samples
have a similar distribution (Fig.\ 1). Plotting Li against metallicity in stars with and
without planets, we found a large scatter. Our Fig.\ 2  shows no 
clear dependence on metallicity. Yet this can be hidden by the mass-related depletion.
In fact, we observe old solar-type stars with
metallicities 2--3 times solar and with abundance of Li similar to the
Sun. This suggests that the metallicity is not the key parameter
determining the  Li abundance in these stars (Pasquini et al.\ 1994). On the other
hand, our plot of Li against $T_\mathrm{eff}$ for the stars of both samples
(Fig.\ 3) does not show anything peculiar. Except for a few stars
 occupying a small  area between 1.0 $< \log \epsilon(\rm Li) <$ 2.2 and
5300 K $< T_\mathrm{eff} < $ 5700 K, this morphology is not different from that
observed in the field stars. However, the low number of stars in the comparison sample with
detectable  Li in their atmospheres (Table 2) does not allow us to
arrive at any firm conclusions.

\subsection{Comparison sample of Chen et al.\ (2001)}

To make this comparison possible we have decided
to use data from the literature. Lithium abundances in field stars from
Chen et al.\ (2001) were used to compare stars with and without exoplanets.
We have removed four stars with exoplanets from the list of Chen et al.\ and
used their data as a comparison sample of stars without planets. Most of the
targets from Chen et al.\ are bright nearby solar-type stars which are
part of various radial velocity surveys. Therefore, it is very unlikely
that the sample contains more stars with exoplanets. Note also that most
of the targets in this sample have solar metallicities  or lower.
Given the strong dependence between the presence of planets and the
metallicity of the parent star (Santos et al.\ 2001, 2003a) we do not expect 
the sample of Chen et al.\ to contain more than one or two so far unknown
planet hosts.

In Table 3, we show the effective temperature distribution of the stars in the
planet host and comparison samples in the temperature range 5600--6350 K.
The planet host sample is biased against lower  and higher
temperatures (Santos et al.\ 2003a); therefore, in what follows, we have not considered such stars
in any of the two samples. The size of the bin used in the table has been chosen taking into
account that the errors in the temperatures are of order 70 K. The three bins represent
three major groups of stars according to the mass in their superficial convective
zones. In the lower temperature bin the mass of the convective zone
is a steep function of temperature, while in the other two bins,
this mass does not change drastically. The third bin is just at a temperature
below the Boesgaard \& Tripicco gap (Boesgaard \& Tripicco 1986).
The table shows that the planet host and the comparison sample have comparable
fractions of stars in each bin.

The lithium distribution in the planet host and Chen et al.\ comparison samples
is shown in Figure 4.  The histogram reveals a marginally statistically significant 
excess of planet
host stars with 1.0 $< \log \epsilon(Li) <$ 1.6. It may be expected that these remarkably
depleted stars come from  the lower temperature bin (deeper superficial zones
and potentially able to sustain a more efficient destruction mechanism).
Looking at Table 3 we see that the planet host sample contains a slightly larger 
relative number of stars in this bin, which would favour a slightly larger
relative fraction of lithium-depleted stars in the planet host sample, but
this is not enough to explain the differences in the histograms of the two
lithium distributions. Measurements of low lithium abundances require high S/N
spectra as the equivalent width of the Li line in these  stars varies between
2 and 8 m\AA.  Our high quality spectra allow a clear detection of the Li line in all cases
with EW $\geq$ 2m\AA. The S/N of spectra of Chen et al.\ (2001) is similar and
therefore we have no reason to suspect that the excess of "Li-poor" stars with
planets is a bias. Thus, we think the effect is real.  In Fig. 5  we find that the lithium
abundances of planet host stars with effective temperatures between 5850 and 6350 K
are similar to those in the Chen et al.\ comparison sample.
While at lower effective temperatures the planet host stars show
on average lower lithium abundances than stars in the comparison sample.
The excess of Li poor planet host stars found in the histogram of Fig. 4
is concentrated in the range 5600 K $< T_\mathrm{eff} < $5850 K.

In Fig.6 we can clearly see that the behaviour of the lithium abundances in the high
 metallicity planet host stars differ only  with respect the comparison sample in the
temperature range  5600 K $< T_\mathrm{eff} < $5850 K. In this temperature range we do
not find a single example of planet host star with high lithium abundance
(i.e.\ log Li $\ge$ 2.0) while in the comparison sample there are many. However,
we should admit that the comparison sample of Chen et al. (2001) does not contain
many stars with [Fe/H] $>$ 0. Future observations of metal-rich stars without
planets (if there are any) may help to confirm our conclusions.

\section{Correlation with orbital parameters}

In Fig.\ 7 we plot the surface abundance of Li against the eccentricities of
planetary orbits. As discussed above, consumption of a slowly migrated inner
planet may increase the surface abundance of Li and modify the  eccentricities of the
remaining planet(s). A similar effect maybe produced if the  ingestion
of a planet is caused by multi-body interactions in the system.
Except for a possible gap at 0.2 $< e <$ 0.4 and 1 $< \log \epsilon({\rm Li}) < $1.6,
our plot does not show any trends.

\begin{figure}
\psfig{figure=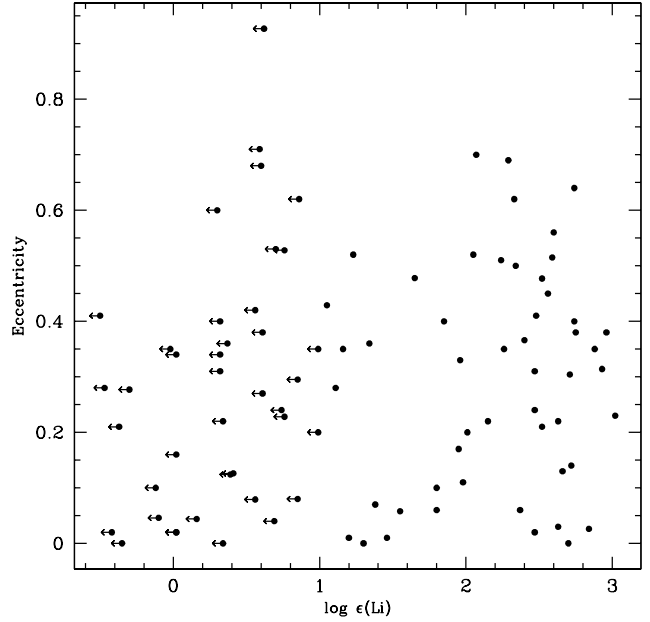,width=9.5cm,height=9.5cm,angle=360}
\caption[]{Eccentricity for the planetary companions against surface abundance of Li.}
\end{figure}

\begin{figure}
\psfig{figure=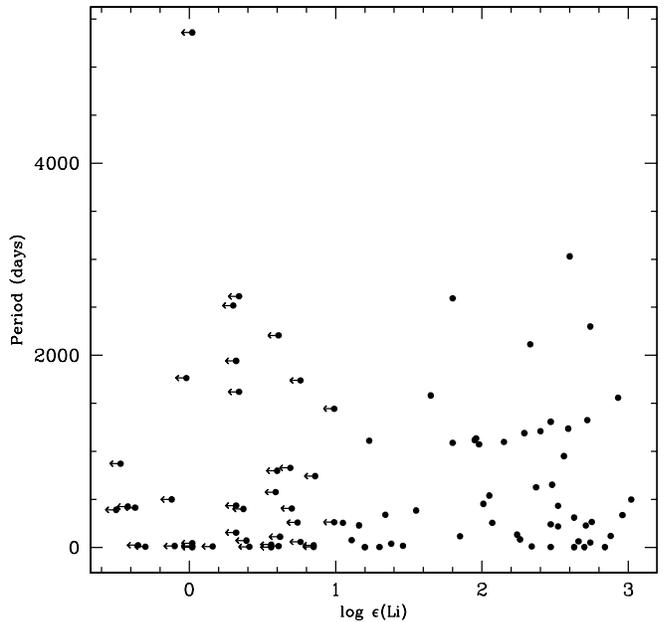,width=9.5cm,height=9.5cm,angle=360}
\caption[]{Orbital period for the planetary companions against surface abundance of Li.}
\end{figure}

\begin{figure}
\psfig{figure=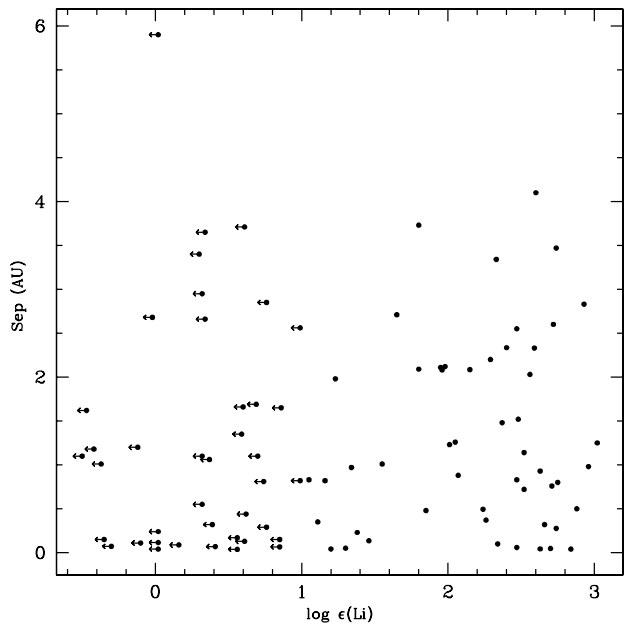,width=9.5cm,height=9.5cm,angle=360}
\caption[]{Separation (semimajor axis) for the planetary companions against abundance of Li.}
\end{figure}

\begin{figure}
\psfig{figure=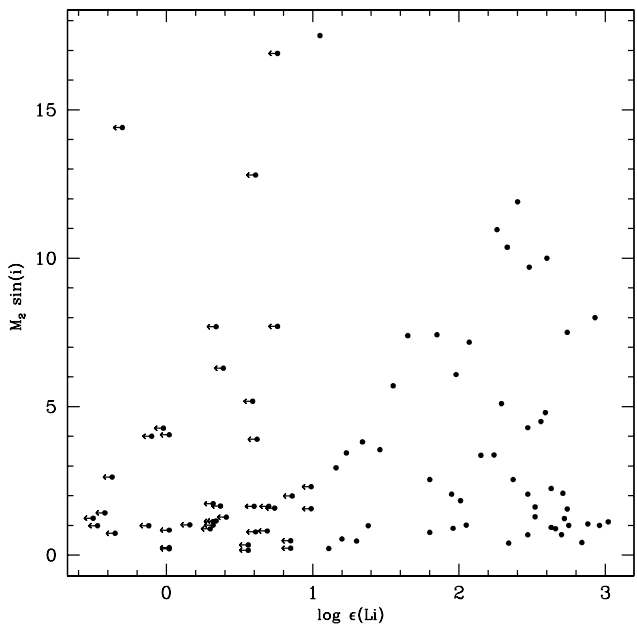,width=9.5cm,height=9.5cm,angle=360}
\caption[]{Minimum mass for the planetary companions against surface abundance of Li.}
\end{figure}

Li abundance against the orbital period is shown in Figure 8. As we can see, there 
seems to be a possible lack of long-period planets orbiting Li-poor stars with
1 $< \log \epsilon({\rm Li}) < $1.6. Apparently, all these planets,
except one, have periods less than 500 days.  All these short-period planets are orbiting
their parent stars at less than 1 AU (Fig.\ 9). Considering stars with 
1.6 $<  \log \epsilon {\rm Li} <$ 3, we find that 50\%  have periods more than 500 days.

In Fig.\ 10 we plot the minimum mass for the planetary companions against Li.
A first look at the plot immediately suggests a lack of massive
planets with $M >$ 4 $M_{\rm J}$ orbiting "Li-poor" stars with
1 $< \log \epsilon({\rm Li}) < $1.6. The only exception is
HD\,202206, which hosts a brown dwarf with a mass 17.5$M_{\rm J}$. When considering
stars with $\log \epsilon({\rm Li}$) between 1.6 and 3, we find that
about half  host planets with  $M>$ 4 $M_{\rm J}$. Obviously, there is a
link between  Mass--Li and Period--Li relationships. This may be associated with the already
proposed correlation between mass and period of planetary
 companions  (Zucker \& Mazeh 2002; Udry et al.\ 2002).
It would be interesting to get higher quality Li measurements for
stars with massive planets and investigate  whether there are no  long-period
massive planets around Li-poor
($1 <\log \epsilon {\rm Li} \leq $ 1.6) exoplanet hosts.

\section{Discussion and Conclusions}

It has been proposed that stars with short-period planets have higher
metallicities among  planet hosts (Queloz et al.\ 2000). If confirmed,
this fact could be interpreted in several ways. For example, inward migration
can produce a metallicity excess (Murray et al.\ 1998) because of the accretion of
planetesimals. One can also imagine that the formation of inner planets is favoured by
the metallicity. Recently Santos et al.\ (2003a) found an indication (which, however, is
statistically not significant) that low mass planets  mostly orbit metal-poor stars
(e.g.\ Udry et al.\ 2002). Apparently, planet host stars more frequently show Li 
abundances in the range
log $N$(Li) = 1.0--1.6 than field stars. These abundances occur in stars
with effective temperatures between 5600 and 5850 K, where we expect well
developed convective zones and a significant depletion of Li. Are planet host stars in this
temperature range more efficient at depleting lithium than single stars?
What is the reason for their different behaviour in comparison with stars without
planets? Why there are no significant differences with field stars
in other Li abundance ranges?

Many processes discussed in this article may modify the surface abundance of Li in stars
with exoplanets. By what amount and when depends on many parameters involved in the 
complex star--planet
interaction. Given the depth of the convection zone, we expect that any effects on the Li
abundance will be more apparent in solar-type stars. Lower mass stars have deeper convective
zones and destroy lithium  very efficiently, so we frequently only set upper limits to the
abundance, which makes it difficult to find correlations with any parameter
affecting Li abundance. On the other hand,
the convective layers of stars more massive than the Sun are too far to reach
the lithium burning layer. These stars generally  preserve a significant fraction of
their original lithium. The relatively small dispersion of lithium abundances in these
hotter stars is clearly seen in Figure 6. It is thus also more difficult to detect any external
effects on the surface lithium.

Solar-type stars are possibly the best targets for investigating any possible effect of planets
 on the evolution of the stellar atmospheric Li abundance.
In these stars we find a lower average Li abundance in planet host stars
than in the field comparison sample (Fig.\ 6, lower panel). 
There are at least two possible hypothesis for the lower Li abundance in exoplanet hosts.
It is possible that proto-planetary disks lock a large amount of angular momentum and
therefore create some rotational breaking in the host stars during their pre-main sequence
evolution. The lithium is efficiently destroyed during this process due to an increased mixing.
The apparent extra depletion may be also associated with a planet migration mechanism at early
times in the evolution of the star when the superficial convective layers may have been 
rotationally decoupled from the interior. Strong depletion may be caused by an effective
mixing caused by Migration-triggered  tidal Forces, which create a shear instability.
The mass of the decoupled convection zone in these stars is comparable to the masses of 
the known Exoplanets; therefore, the migration of one or more planets could indeed
produce an observable effect. The migration of planets may also produce
the accretion of protoplanetary material and/or planets, inducing
metallicity enhancement, and some fresh lithium could also be incorporated
in the convective zone. However, if this takes place in the early evolution
of the star, this lithium will most probably be destroyed.

Our observations suggest that Li abundances in stars with short-period planets may be
influenced by the presence of planets. More observations would be welcome to tackle this problem.

\def\baselinestretch{1}

\begin{table}
\caption[]{Distribution of stars in the comparison samples of Chen et al.\ (2001)
and planet hosts}
\begin{tabular}{lcccc}
\hline
\noalign{\smallskip}
Range & Planet hosts & Comparison sample  \\
\hline \\
5600 $< T_\mathrm{eff} \leq$ 5850     &  39\% (22)  & 32\% (50)   \\
5850 $\leq T_\mathrm{eff} < $ 6100    &  34\% (19)  & 40\% (64)  \\
6100 $\leq T_\mathrm{eff} \leq$  6350 &  27\% (15)  & 28\% (43)  \\
\noalign{\smallskip}
\hline
\end{tabular}
\end{table}

\end{document}